12 Oct, 2011

Arend Nijhuis
University of Twente
Netherlands
a.nijhuis@utwente.nl

## Optimisation of ITER Nb$_3$Sn CICCs for coupling loss, transverse electromagnetic load and axial thermal contraction

A. Nijhuis, E.P.A. van Lanen, G. Rolando

**Abstract**

The ITER cable-in-conduit conductors (CICCs) are built up from sub-cable bundles, wound in different stages, which are twisted to counter coupling loss caused by time-changing external magnet fields. The selection of the twist pitch lengths has major implications for the performance of the cable in the case of strain sensitive superconductors, i.e. Nb$_3$Sn, as the electromagnetic and thermal contraction loads are large but also for the heat load from the AC coupling loss. At present this is a great challenge for the ITER Central Solenoid (CS) CICCs and the solution presented here could be a breakthrough for not only the ITER-CS but also for CICC application in general. After proposing longer twist pitches in 2006 and successful confirmation by short sample tests later on, the ITER Toroidal Field (TF) conductor cable pattern was improved. As the restrictions for coupling loss are more demanding for the CS conductors than for the TF conductors, it was believed that longer pitches would not be applicable for the conductors in the CS coils. In this paper we explain how with the use of the TEMLOP model and the newly developed models JackPot-ACDC and CORD, the design of a CICC can be improved appreciably, particularly for the CS conductor layout. For the first time a large improvement is predicted not only providing very low sensitivity to electromagnetic load and thermal axial cable stress variations but at the same time much lower AC coupling loss.

Reduction of the transverse load and warm-up cool-down degradation can be reached by applying longer twist pitches in a particular sequence for the sub-stages, offering a large cable transverse stiffness, adequate axial flexibility and maximum allowed lateral strand support. Analysis of short sample (TF conductor) data reveals that increasing the twist pitch can lead to a gain of the effective axial compressive strain of more than 0.3 % with practically no degradation from bending. This is probably explained by the distinct difference in mechanical response of the cable during axial contraction for short and long pitches. For short pitches periodic bending in different directions with relatively short wavelength is imposed because of lack of sufficient lateral restraint of radial pressure. This can lead to high bending strain and eventually buckling. Whereas for cables with long twist pitches the strands are only able to react as coherent bundles, being tightly supported by the surrounding strands, providing sufficient lateral restraint of radial pressure in combination with enough slippage to avoid single strand bending along detrimental short wavelengths. Experimental evidence of good performance was already provided with the test of the long pitch TFPRO2-OST2, which is still until today, the best ITER type cable to strand performance ever without any cyclic load (electromagnetic and thermal contraction) degradation.

For reduction of the coupling loss, specific choices of the cabling twist sequence are needed with the aim to minimize the area of linked strands and bundles that are coupled and form loops with the applied changing magnetic field, instead of simply avoiding longer pitches. In addition we recommend increasing the wrap coverage of the CS conductor from 50 % to at least 70 %. A larger wrap coverage fraction enhances the overall strand bundle lateral restraint.

The long pitch design seems the best solution to optimize the ITER CS conductor within the given restrictions of the present coil design envelope, only allowing marginal changes. The models predict significant improvement against strain sensitivity and substantial decrease of the AC coupling loss in Nb$_3$Sn CICCs, but also for NbTi CICCs minimization of the coupling loss can be achieved. Although the success of long pitches to transverse load degradation was already demonstrated, the prediction of the combination with low coupling loss needs to be validated by a short sample test.

## 1 Introduction

Cable-in-conduit conductors (CICCs) for the ITER Central Solenoid (CS) coils are consisting of superconducting strands and copper strands with a diameter of 0.83 mm that are cabled by twisting in several stages, thus creating a wavy pattern of the strands throughout the cable. The bundles are twisted to counter coupling losses caused by induction currents due to time-changing external magnet fields [1]. The conductors have a void fraction ($vf$) of about 30 % and the strand bundle is cooled by helium flowing in the cable voids and is contained in a steel jacket. When cooling from reaction heat treatment to cryogenic magnet operation temperature, the differential thermal contraction between conduit material and strand bundle causes besides a contraction of the strands in axial direction, bending of the strands although this is strongly connected with the selected cable twist pattern. This axial and bending strain is again to some extent moderated by the coil hoop stress during magnet operation. The CS conductors are carrying more than 40 kA in a magnetic field locally exceeding 12 T, hence subjecting them to severe transverse loading due to the Lorentz forces. This imposes distributed magnetic loads along each strand, but also cumulated loads from other strands transferred by the strand-to-strand contacts [2] [3] [4]. These bending and contact loads on the $Nb_3Sn$ strands create a periodic strain variation along the filaments. The magnitude and periodicity of the periodic strain pattern in combination with the ability of a strand to redistribute the current between the filaments determines the impact on the critical current ($I_c$) and steepness of the voltage current transition ($n$-value).

Since the first tests of the Central Solenoid Model Coils (CSMC) in Japan, great effort has been spent to the understanding of the unexpected severe degradation of the conductors compared to that of the single strand performance [5], [6], [7], [8], [9], [10], [11]. Many models were presented describing the degradation based on sometimes- cruel current non-uniformity to often-severe strand bending but not coming with adequate solutions although it was predicted that a lower void fraction and shorter cabling pitches could perhaps partly confine the transverse load degradation [12]. In the meantime, the ITER conductor design has already been modified compared to the CSMC layout. It was assumed that the resistance to the degradation could be increased by using a steel jacket, providing additional thermal pre-compression in reducing the tensile strain levels being associated with strand bending and filament cracking [2], [3], [4], [5] [12], [13]. The void fraction was reduced from 36 % to 33 %, the Cu:nonCu ratio of the strands was reduced from 1.5 to 1.0 with introduction of one copper strand in the first triplet and the non-copper material in the cross section was increased by 25%. In addition, newly developed, so-called high $J_c$ advanced $Nb_3Sn$ strands were being pursued in attempt to compensate for the performance loss. Still the short sample tests did not lead to the desired performance [14][15].

However, until then it was never investigated whether the selection of the twist pitch lengths could also have noteworthy implications on the performance of the cable in the case of strain sensitive superconductors, i.e. $Nb_3Sn$, as it strongly affects the overall cable stiffness. The critical current and temperature margin is not only affected by the thermal contraction of the composite materials, but also largely by electromagnetic forces as the drawback of such cables is that the strand magnetic loads are not well supported [2]-[4]. A combined effort of dedicated modeling and experiments on strands and cables was required to better understand the involved interactions between strands and applied loads. The influence of various loads and deformations (axially tensile, bending strain, contact load from crossing strands or homogeneous transverse load) that occur in a CICC, were extensively studied with different probes in the TARSIS facility at the University of Twente [16][17][18][19]. The key advantage of the TARSIS setup is its ability to measure both the amplitude of deflection or deformation and the applied force up to thousands of load cycles with high precision, enabling a full axial and transverse stress-strain analysis. The specific cable data on AC loss, interstrand resistance distribution and mechanical stiffness were obtained from Twente Cable Press experiments [20] [21]. Since it is the stress distribution, - originating from differential thermal contraction and electromagnetic load - that drives the evolution of the strain distribution with cyclic loading, the stiffness of the strand and the role of strand support were definitely identified as key parameters. The axial tensile and compressive strain parameterization, identifying its sensitivity for applied axial strain, magnet field and temperature were performed with the Pacman spring [22].

The efforts lead to the innovative prediction with the TEMLOP model first presented in 2006 [23]. The TEMLOP model describes the mechanical response and interaction of individual strands within a cable bundle and associated change in transport properties, clearly showing that applying longer cabling pitches provides a convenient solution against transverse load degradation. Insufficient practical experience with varying cabling pitches and a consequential conservative approach, explains why this factor remained hidden for so long. It was assumed until then that only shorter cabling pitches could lead to a better performance obviously by shortening the bending beam [12]. This was indeed confirmed by the TEMLOP model, however, we will explain further on in the paper why the application of short pitches may be advantageous for transverse load but does not lead to the desired stable performance in view of thermal axial cable contraction.

The predicted relationship between transverse load and the cable twist pattern i.e. the tendency for $I_c$ reduction and peak bending strain as a function of the characteristic bending wavelength is depicted in Figure 1 [23]. The characteristic bending wavelength is the average distance between strand to strand contacts that act as supports and transfer the load from strand to strand through the layers in the cable.

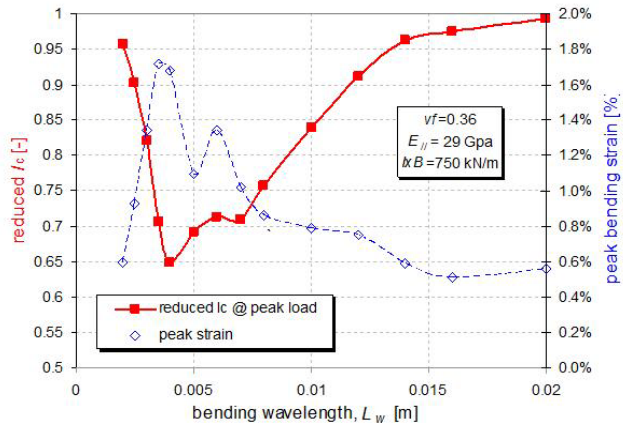

*Figure 1. The TEMLOP prediction of Ic reduction and peak bending strain as a function of the characteristic bending wavelength as described in [23].*

This predicted relationship was successfully verified during two experimental campaigns in SULTAN with the TFPRO2-OST1 (TF baseline Option1 cable pattern) and OST2 (TEMLOP long pitches) in 2007 [24][25][26] and the so-called PITSAM #3 and #5 samples in 2008 [27]. The twist pitch schemes of the TFPRO2 conductors are listed in Table 1.

Table 1. Cabling layout parameters for TF and CS variations.

|  | Previous TF reference Option 1: TFPRO2-OST1 | TEMLOP Long Pitch: TFPRO2-OST2 | Final Reference TF Option 2 | Baseline CS conductor (PA) CSJA-01 | CS alternative short Lp [28] | CS alternative CS-Insert based [28] |
|---|---|---|---|---|---|---|
| Cable pattern | (2sc + 1Cu) x 3 x 5 x 5 +3x4Cu x 6 | | | (2sc + 1Cu) x 3 x 4 x 4 x 6 | | (3sc x 3 x 3 x 5 + core) x 6 core 3 x 4 |
| Twist pitches (mm) |  |  |  |  |  |  |
| Stage 1 | 45 | 117 | 80 | 45 | 20 | 45 |
| Stage 2 | 87 | 182 | 140 | 85 | 45 | 85 |
| Stage 3 | 126 | 245 | 190 | 145 | 80 | 145 |
| Stage 4 | 245 | 415 | 300 | 250 | 150 | 250 |
| Stage 5 | 460 | 440 | 420 | 450 | 450 | 450 |
| Nb$_3$Sn strand Cu-to-non-Cu ratio | 1.0 | 1.0 | 1.0 | 1.0 | 1.0 | 1.5 |
| Number of sc strands | 900 | 900 | 900 | 576 | 576 | 810 |
| Void fraction (annulus) [%] | 29 | 28 | 30 | 33.6 | 30.6 | 31.4 |

The PITSAM conductors with 30 % void fraction were cabled as 3x3x3x4 with 48 superconducting strands and following 58, 95, 139, 213 mm (PITSAM2) and 34, 95, 139, 213 mm (PITSAM5-short) and 83, 140, 192, 213 mm

(PITSAM5-long) cable twist pitches. The twist pitch scheme of the TFPRO2-OST2 was based on the TEMLOP prediction aiming for a characteristic bending wavelength of more than 20 mm, while the ITER reference cable pattern used until then had a wavelength estimated below 10 mm. The experimental data in terms of current sharing temperature ($T_{cs}$) reduction versus the twist pitch of the first cabling stage is presented in Figure 2. The short pitches of the TFPRO2-OST1 results into a large degradation of the $T_{cs}$ and further degradation with cycling. While the long pitches of the TFPRO2-OST2 seem to experience no degradation at all with cycling and remains close to the strand performance.

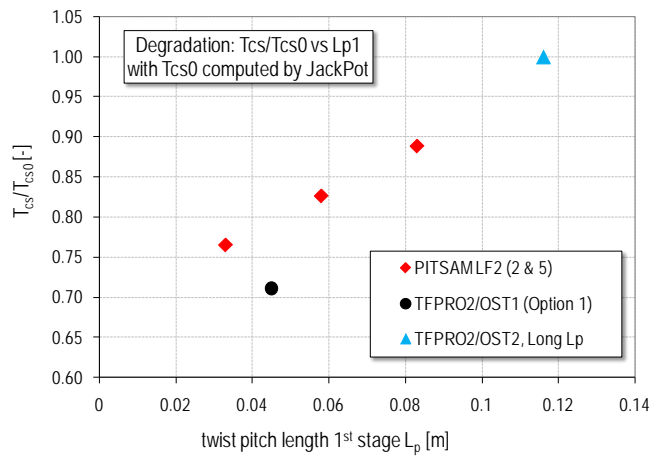

*Figure 2. Tcs reduction as a function of $1^{st}$ stage cabling pitch for the TFPRO2 OST1 and OST2 and the PITSAM 2 and 5 conductors.*

The PITSAM CICCs with different cabling patterns show similar tendency with smaller degradation due to lower transverse "BxI" load as the number of strand layers in the cable is much less than for the TFPRO2. The reduction in $T_{cs}$ is computed with the JackPot model [29][30][31][32], taking into account all strand trajectories, the strand magnet field, temperature and strain scaling, the applied and cable self-field, to assess the maximum possible virgin $T_{cs}$ based on the strand properties and compare it to the SULTAN data. In addition another series of sub-size CICCs, also contained a variation of the twist pitch with respect to the traditional cable pitch sequence of 45x85x125 mm but in the direction of shorter pitch lengths 35x65x110 mm [33]. The results clearly showed a worse performance for the cable with shorter pitches compared to the standard ones. Despite the successful test result, the TFPRO2 long pitch cable layout was not fully adopted but a twist scheme with somewhat shorter pitches was eventually decided for the final ITER TF conductor design, referred to as Option 2 (see Table 1) [34]. Although not sufficient elementary analysis was available at that time, the motivation for this compromise was partly based on difficulty expected from cable manufacture and having concern about high interstrand coupling loss. Favorably, since 2008 a large series of TF conductors have been qualified reaching the required Tcs criterion [35] in strong contrast to earlier results [15][26]. However, it should be noted that due to the mentioned compromise, the present TF cable design is not fully optimized with respect to transverse electromagnetic load but only adapted to avoid abundant filament damage and so still most TF samples show degraded performance with cycling. Therefore, not all ITER TF Option 1 conductors perform below the ones with Option 2 cabling pattern and some scattering is observed [36], [37], [38]. It is believed that this can be attributed to the many aspects that are involved with conductor manufacture, sample preparation and testing, e.g. joint properties, local cabling variations, jacket material properties or untwisting effects during cable insertion [39]. This already leads to somewhat unexpected dissimilar results even in case of "similar" samples [35], [36], [37], [38]. Nonetheless, the convincing results of the direct comparative studies provide sufficient confidence to take the long twist pitch cabling pattern of the TFPRO2-OST2 as the starting point of our CS conductor optimisation. The TFPRO2-OST2 sample based on the TEMLOP design is still until today the best performing TF conductor ever, in spite of the used strand with relatively inferior properties [25]. The heat load associated with applied magnetic field ramps, playing just a marginal role for the TF conductors, must be well controlled in the case of CS conductors [40]. For this reason, an optimization with respect to the generation of interstrand coupling loss is an important design aspect and is adopted as one of the main objectives of this paper. The eventual requirements set on the cabling pattern in relation to coupling loss optimization, calls for clear conditions with regard to the relevant frequency (dB/dt) spectrum of the applied changing magnet fields (and the allowed heat load on the conductor). The use of a single coupling loss time constant for such computations leads to deceptive results and a conceptive formulation as presented in [41][42] is not easy to scale on an accurate way. The only way out is the utilisation of a dedicated cable model incorporating all strand trajectories, inductive coupling and associated current distribution and so for this purpose we developed JackPot-ACDC. The broadly presupposed disadvantage of longer pitches leading to higher coupling loss is carefully investigated with the model. The overall object of this paper is to find an optimized design of a CICC cable layout for electromagnetic and thermal load degradation but with minimum heat load from interstrand coupling loss, with particular reference to the ITER CS coil conductors.

## 2 Degradation of $T_{cs}$ with cycling

In this section we provide an overview of our interpretation of the phenomena involved with performance degradation of ITER Nb$_3$Sn conductors. The present baseline ITER CS cable twist pattern is actually similar to the previous TF Option 1 design (see Table 1), although the number of strands differs for the higher cable stages. For a first order estimation of the transverse load comparison between TF and CS layout, often the peak current and magnet field are taken and multiplied ($IxB$), I=68 kA, B=11.8 T for TF and for CS two operating conditions can be considered: I=40 kA, B=12.34 T and I=45.4 kA, B=11.78 T. When only comparing the $IxB$, we find 800 kAT for TF and 500 kAT for the CS, giving a ratio of 0.6 and leading to the impression that the transverse load is much less severe for the CS. In reality the strand current is practically similar for both CS and TF conductors, this means that the number of superconducting strand layers accumulating in the load direction is essential. For TF this amounts to 900^0.5 (30 layers) by approximation and for CS this is 576^0.5 (24 layers), leading to a ratio of 0.8, illustrating that the actual transverse load for CS compared to TF is not so much different in reality. This is also illustrated by TEMLOP computations. In [23] it was explained that apart from the electromagnetic load and strand mechanical properties, the associated bending deflection is also determined by the distance between the supporting strands. This characteristic bending wavelength ($L_w$) is in turn determined by the cabling pattern of the strand bundle and is connected to the number of elements per sub-cable and the twist pitches and finally by the load. With TEMLOP [23] we computed the transverse load $I_c$ degradation for the present TF layout (Option 2, 30 % void fraction) and the ITER baseline CS layout (33 % void fraction) for similar strand properties. The result is depicted in Figure 3, showing a larger degradation for the present CS layout due to shorter pitches and higher void fraction. The other curves in Figure 3 will be discussed further on.

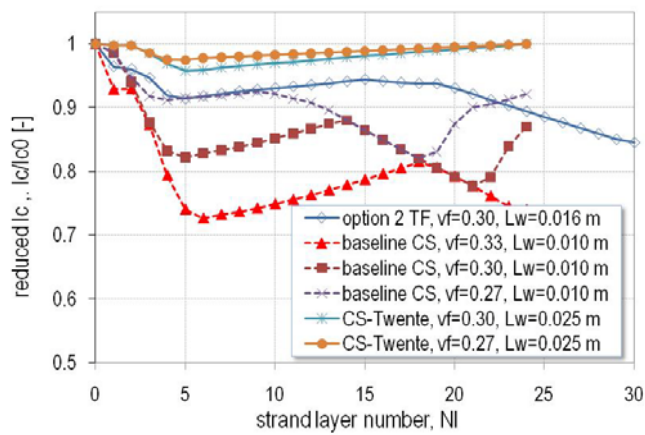

*Figure 3. TEMLOP analysis of the transverse load effect for TF-Option2 and various CS cable layout options for similar strand (OST-II type [25]). The influence of the void fraction is shown in relation to the cable layout (characteristic bending wavelength Lw).*

The main difference, apart from the peak $IxB$ load, concerns the difference in twist pitches and void fraction (see Table 1).

Recently (2010) the first CS type of conductor (CSJA-01) with ITER baseline layout was tested in the SULTAN facility. The CSJA-01 sample was manufactured by JAEA (Japan), following non-standard procedures and the sample manufacturing details are still under review [43], [44]. It came out that the CSJA-01 exhibited a continuous degradation of the $T_{cs}$ with electromagnetic (EM) and thermal warm-up cool-down (WUCD) cycling, after showing an increase at initial load cycles. The average overall decrease of the $T_{cs}$ after 6,000 EM loading cycles and a WUCD cycle amounts to 1 K, without any sign of stabilization [43], [44]. This represents a global degradation of almost 0.2 mK/cycle. Also the second sample, CSJA-02, being still under test, shows a continuous degradation of up to 0.7 K after 10,000 cycles for both legs without showing saturation (0.07 mK/cycle). The decrease of the $T_{cs}$ with cycling for the CSJA-01 up to the first 1000 cycles followed more or less the tendency as observed for the TF samples. Except that the CSJA-01 exhibits an initial $T_{cs}$ increase during the first cycles, that is not observed on most TF samples. The typical evolution of the $T_{cs}$ along load cycling in a SULTAN short sample test is a fast initial decrease, followed by a more gradual decay and a sudden decrease after a WUCD cycle. It should be noted that the degradation of the CSJA-01 seems somewhat extreme, however, it is not that only Sultan CICC samples suffer from transverse load and thermal cycling [35], as the CSMC inner layer (conductor 1A) performance appears to be slowly but steadily degrading with time with a loss in $T_{cs}$ of 0.3 K to 0.4 K and decrease of $n$-value from 9 to 5, after limited number of charging cycles (160) and only few warm up – cool down (WUCD) cycles with respect to the first test campaign in 2000, [5], [6], [7], [8], [45]. The overall degradation rate for the CSJA-01 amounts roughly to 0.2 mK/cycle but the CSMC-1A rate is even 2.5 mK/cycle although this is mostly associated with WUCD cycles [45]. We should remark here that defining a rate of $T_{cs}$ degradation per load cycle depends highly on the circumstances, is not a well defined parameter but can be seen as a rough indicator for degradation. The CS-Insert coil performed 10,000 cycles with an overall degradation of 0.05 mK/cycle, with the main part of the degradation during the first 1000 cycles [5]. It was observed that the largest degradation of the CS-Insert occurred during quenches of the coils. However, it is not unlikely that a quench may cause mechanical effects in a cable that are comparable to the thermal load of a WUCD cycle, i.e. strand re-arrangement and slippage as discussed below. In that sense we should take it at least as an important caveat.

Apart from the $T_{cs}$ degradation as measured at full electromagnetic load cycling, there is also a strong $I{\times}B$ dependency of the performance observed on not only the TF and CS Sultan samples but also on the CS and TF Model and Insert Coils [7], [12], [46], [47].

For the interpretation of CICC behavior in coils and short sample tests we can distinguish two effects connected to strain changes in the Nb$_3$Sn layers. One is the transverse electromagnetic load effect leading predominantly to bending and pinching effects on the strands and the second is the warm up cool down (WUCD) effect, connected to the difference in axial thermal contraction of the cable and its jacket. The cable is in compression and the jacket is in axial tension after reaction heat treatment at 650 C and cool-down to 4.5 K.

As far as the transverse electromagnetic load effect concerns, this is simulated in the Twente Press experiment and the results obtained on three different TF samples (all similar cable pattern) are plotted in Figure 4, [48], [49], [50]. The strands have different stiffness and this seems to be reflected in the cable transverse modulus.

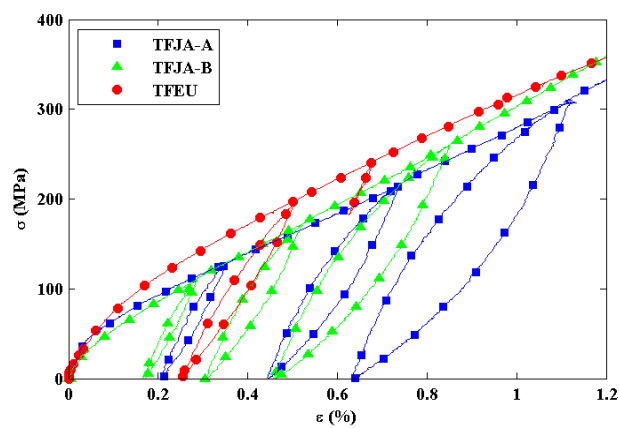

*Figure 4. The axial tensile stress-strain curves of strands from the TFEU3 and TFJA5 samples.*

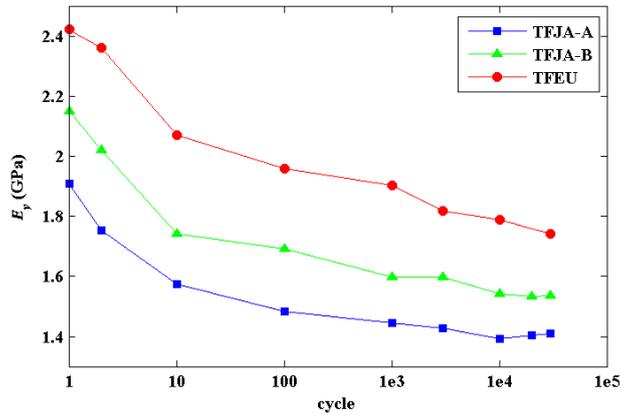

*Figure 5. The transverse cable modulus as a function of number of load cycles from the TFEU3 and TFJA5 samples.*

For all cable samples we observe a progressive reduction of the cable transverse modulus with number of load cycles and an increase of the overall cable compression leading to lower void fraction and a gap between inner jacket wall and cable perimeter (see Figure 6 and Figure 7). The overall cable compression for a CS1 model coil CICC after cycling with full load is 1.1 mm for a void fraction of 36 % [50]. The overall compression for the TF Option 2 CICCs, with a void fraction of 30 %, tested in the press is relatively large and ranges from 1.3 mm to 1.6 mm [48], [49], [50]. The higher compaction of the cable causes disengagement with the jacket due to reduced friction. The cyclic load causes increased strand bending, pinching and yielding of the copper which reduces the effective stiffness of the composite cable. This behavior can be associated with the reduction in $T_{cs}$ with electromagnetic load cycling.

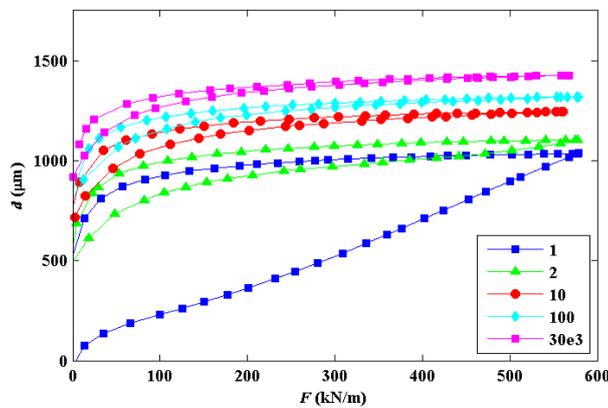

*Figure 6. The stress strain curves of the of the TFJA5-B sample for different number of load cycles, measured in the Twente Press, showing a subsequent increase of the cable deformation with cycling.*

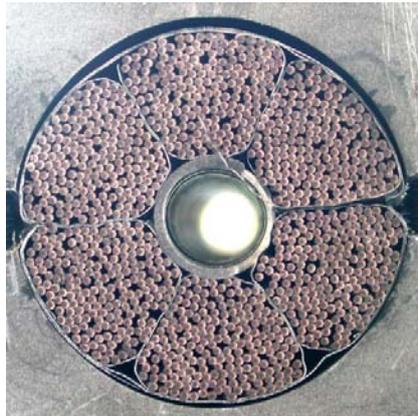

*Figure 7. Cross section of the ITER CS1 Model Coil conductor after 40,000 cycles at peak load. The gap at the top between cable and jacket shows the degree of plastic deformation [50].*

In the Twente Press the load is uniformly distributed along the entire length of the sample, while in a coil gradual variations occur. For a Sultan conductor sample of 3.5 m length, the high field region of 11 T is less than 0.5 m and outside the high field zone a strong field gradient exists. Measurements with strain gauges attached to SULTAN samples revealed that the strain on the jacket reduces in the high field section, while it remains practically constant in the low magnet field sections [54]. A local relaxation of the jacket tensile strain is associated with a reduction in length of the high field section and a local increase of the cable compressive strain, which thereby becomes non-uniform along the sample axis.

The importance of the cable's axial contraction was well recognized and modeling efforts were presented in [3], [55], [56]. In [57], the mechanisms that could create a non-uniform longitudinal strain in the cable of a Sultan sample have been modeled. The first mechanism is associated with a local reduction of the cable effective modulus as a result of the application of the transverse magnetic loads. The second is associated with an extra local plastic deformation of the cable, also as a result of extra strand bending and copper yielding. The degree of degradation strongly depends on the cable stiffness and available strand support. Only when strands have sufficient space available, for example in a cable with an enhanced wavy pattern (short pitches), or high void fraction and in the region of the gap between inner jacket wall and transverse compressed cable, enhanced degradation can take place by severe bending.

For a cable with low coverage of the petal wraps, this may eventually lead to buckling of a (small) fraction of the strands on the cable surface [58]. When resulting in sharp kinks, strands may fail after repeated cycling. Models have shown that groups of saw-tooth buckles are separated by straight slip lengths in ropes representative for multistrand composites like CICCs [59]. The forces involved are the compressive axial load, the frictional resistance to axial slip, and the lateral restraint of radial pressure. The radial pressure is at minimum in the gap region, especially for low petal wrap coverage percentage, high void fraction and short to intermediate twist pitches. The modulus and the friction coefficient have a role in determining the displacement in the slip zone, which affects the compressive axial stress, but dominant properties are the strand bending stiffness (or restraint), which sets the pattern of initial buckling, and the bending-yield moment, which determines the formation of plastic hinges.

For improvement of the CS conductor design, it is obviously recommended to have strands with high bending stiffness [23] but most of all lateral restraint of radial pressure is required to prevent severe bending and buckling. This can be achieved by modifying the twist pattern, searching for maximum strand support and at the same time increase the petal wrap coverage.

Altogether it seems that a primary condition to avoid serious degradation of the conductor during the life time of ITER due to the combined transverse load, axial stress variations and WUCD effects, is significant improvement of the lateral restraint in radial strand support. This is further analysed in the following sections.

### 3 Axial stiffness, cable patterns and void fraction

A mechanical model for a superconducting cable (CORD) was build, which can predict the strain and stress states of all single strands including interstrand contact force and the associated deformation up to three cable stages in a tensile stress-strain test [60][61]. The measured stress strain curves at 4.2 K of ITER superconducting $Nb_3Sn$ and copper strands are used as a direct input for the model. The result in terms of peak bending strain indicate that for every case the bending strain reduces for longer twist pitches. It seems that in particular the shortest twist pitch, which in general is the first stage triplet, dominates the peak bending strain that is attained.

Although for a $Nb_3Sn$ CICC in reality the jacket is in tension and the cable is in compression, these numerical results reflect the bending strain for an axial tensile strain test on a cable. However, the distribution of the contact points and freedom to deflect is considered to be relevant for both, thermal axial compression and transverse electromagnetic force. As already explained in the previous section, it is understandable that the contact point distribution and the lateral support dominate the freedom of spatial periodic strand deflection. In that sense the obtained results may scale with varying pitch combinations under conductor operation performance. This emphasizes that besides low bending strain, we should aim for maximum lateral constraint in radial direction.

There are only very few full-size $Nb_3Sn$ conductor samples available to evaluate the impact of the cable pattern on the axial cable compression. A series of samples is manufactured with the TF Option 2 cable layout, but only few are available with Option 1 and only one with the TEMLOP long pitches layout (TFPRO2-OST2). The DC transport properties of ITER TF conductor samples is measured in the SULTAN facility, but the data processing of the associated current sharing temperature ($T_{cs}$) can sometimes give scope for different interpretation. To extract the conductor's pure performance during such short-sample tests requires a detailed quantitative analysis. This is done with JackPot-ACDC, for which practically all input parameters are based on experimentally verified data. The TFPRO2 and TFJA3 samples, which have different cabling and joint layout have been analysed extensively [62]. Thanks to the detailed characterization of the joints by measurement of interstrand resistances, the interference from the joints could be distinguished from the actual cable performance [63].

For these $Nb_3Sn$ conductors, the effective axial strain is the only free parameter left in the model for matching the simulations with the SULTAN $T_{cs}$ tests and the results are summarized in Table 2. The calculated effective axial strain of the $Nb_3Sn$ layers is plotted against the shortest (first) stage twist pitch in Figure 8.

Table 2. Effective axial compressive strain in the $Nb_3Sn$ layers for three different cable patterns but practically similar void fraction [62].

| Sample ID | Cabling | $\varepsilon_{axial}$ in simulation |
|---|---|---|
| TFPRO2 OST1 | TF Option 1 | -0.69 % |
| TFPRO2 OST2 | TEMLOP Long Lp | -0.36 % |
| TFJA3-J | TF Option 2 | -0.60 % |
| TFJA3-F | TF Option 2 | -0.58 % |

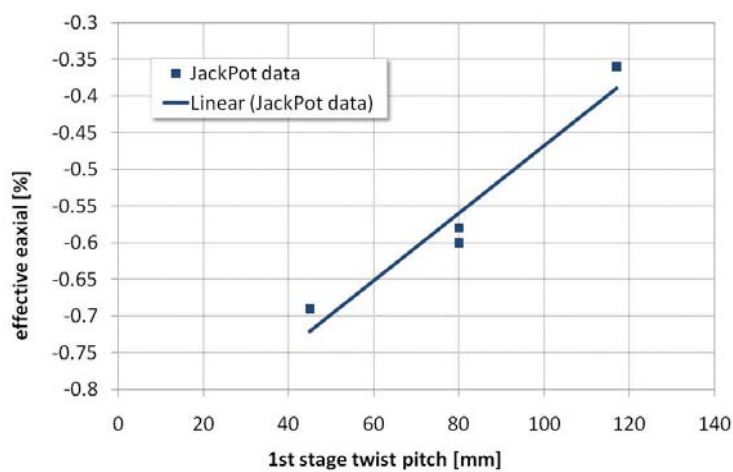

Figure 8. Effective axial strain of three different cable patterns as a function of $1^{st}$ stage cabling pitch [62].

It appears that the effective axial strain in the four analyzed conductor samples becomes less compressive with increasing twist pitches while at the same time their relative performance improves appreciably. Although the correlation from Figure 8 provides a clear direction for our overall conductor analyses, the effective axial strain cannot be distinctly separated from transverse load influence. Looking at other TF samples with option 2 cable pattern, being not analyzed in combination with extensive joint resistance distribution measurements and obtained through the use of a different electromagnetic code as reported in [35], the average of the effective axial strain values were found equal to -0.69% before cycling and to -0.73% after cycling. On first sight, these values seem not support the tendency from Figure 8 obtained on just a few samples. As mentioned previously, a spread may be anticipated regarding all aspects involved with conductor manufacture, sample preparation and testing [39]. That means that the correlation in reality is less pronounced as presented in Figure 8. On the other hand, we refer to the analsysis presented in [37] and [36], confirming by statistical approach a better performance for Option 2 compared to Option 1 cabling, as from this work it was concluded that 45 mm triplet twist pitch and so-called negative VI slopes are often found in underperforming conductors compared to 80 mm twist pitch from Option 2. Looking at the Tcs after the first cycle as a function of the witness strand $I_c$, the author proposes a separation between "normal" degradation and "strong" degradation parallel to the potential $T_{cs}$ performance prediction. In that case more than half of the Option 1 conductors stay below that line, while more than 85% of the Option 2 conductors remain above the line [38]. This statistical approach supports the tendency visualized in Figure 8.

From the previously summarized data and analysis we take that longer twist pitches provide significant higher cable performance even leading to compressive strain in the Nb$_3$Sn layers of up to -0.36 % (for TF CICCs), i.e. a much higher $T_{cs}$ and no cycling degradation.

Although it may be in the error bar of the experiment, it may also not be just by chance that in the linear fit of Figure 8, the cables made of bronze wire (TFJA3) have relatively higher compression than the ones made with internal tin wires (TFPRO2). For a free standing strand, the axial compression of the Nb$_3$Sn filaments is mostly somewhat higher than for internal tin wires due to the difference in thermal contraction coefficient of the composite materials [64][65][66].

Here we can add that the TEMLOP model predicts good performance for short pitches when describing transverse load effects. However, for compressive axial load it is likely that the pitches must be relatively short in order to prevent buckling and the results in Figure 1, Figure 2, and Figure 8 do not leave much scope for fast improvement towards shorter pitches.

To our knowledge there are no data on the overall strain change of the jacket for the TF conductors in Figure 8 so it is not confirmed whether the axial stress is higher in the jacket of the conductor with longer pitches. However, it is likely that the jacket eventually contracts to a practically similar extent but that the mechanical response of the strands in the cable works in a different manner depending on the cable pattern, wrap coverage and void fraction. For short and intermediate twist pitches, the axial contraction of the cable causes bending as the strands already have a wavy pattern, are periodically clamped by other strands, not supported in between and local sliding can even lead to buckling. This is illustrated in Figure 29 (left) where during compressive axial load it is expected that some strands will bend in one direction while other will bend in opposite direction with relatively short wavelength. The strands are sintered after the heat treatment but only after a few loading cycles they become disengaged, which is evidenced by the increase of the interstrand resistance with cycling [20], [42], [48], [49], [50], [51], [52], [53]. The disengagement enhances the opportunity for slippage which then leads to buckling on locations where insufficient lateral restraint of strand deflection is present [58]. Whereas for cables with long twist pitches when subjected to compressive axial load (see Figure 29, right) the strands are being supported by all other strands and there is sufficient lateral restraint of radial deflection in combination with slippage to avoid bending along short wavelengths.

Nevertheless, in any case the total cable length is forced to follow the jacket axial contraction but our conception is that the axial jacket strain is not one-to-one transferred as an axial strain to the individual strands. The cable contraction is likely absorbed by the individual strands as bending along large beam radii and low bending strain combined with minor torsion, which may explain why the effective axial compressive strain of the TFPRO2-OST2 can reach a value as low as -0.36 % in the strands. For a conductor with large twist pitch, the bending strain and torsion appears to remain restricted and degradation hardly occurs. From this we take that the effective axial strain in the strands must be tightly connected to a beneficial combination of cable pattern and void fraction.

The initial thought when selecting the void fraction for a CS improved conductor design could be to take it similar to the one from the TFPRO2-OST2 sample. The influence of the void fraction in relation to the cable layout (characteristic bending wavelength $L_w$) is shown in Figure 3 where the TF Option 2 pattern is compared to several alternative layouts for the CS. For the TEMLOP analysis concerning the effect of transverse load we found that reducing the void fraction indeed improves the performance against transverse load but for $L_w$ reaching the level of TFPRO2-OST (25 mm), the impact of the void fraction seems to become marginal. So for transverse load and long pitches, the role of the void fraction seems to become less critical.

On the other hand, we have observed that during cable manufacture the degree of deformation after the full-size cable compaction step required for jacketing, is less severe for longer pitches compared to short twist pitches. This is illustrated in Figure 9 where two of the recently manufactured CS cable alternatives are depicted.

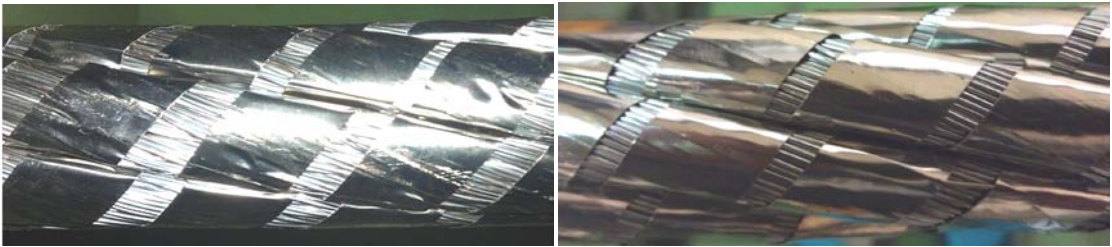

*Figure 9. Two cable sections after compaction to the cable diameter required for jacketing, left the reference CS cable pattern, right the CS-Twente long pitches. The CS Twente design shows significantly less petal deformation. The strongest deformation occurred for the so-called short twist pitch conductor.*

For (very) short twist pitches the petal deformation is severe and the cable is tightly compacted inside the jacket. During the tranverse loading tests in the Twente Press it already appeared that for conductors with similar void fraction and the same load conditions, the deflection is significantly larger (factor of two) for cables with longer pitches (Option 2 compared to Option 1) and as a result leads to much larger gap between jacket and cable [34]. Thus, for shorter pitches there is less scope for deformation of the overall bundle and its petals and so less flexibility. This additional space is likely the reason that more flexibility is provided to the petals to bend along large radii resulting is lower effective axial strain and lower peak bending strains as for long pitches the strands remain subjected to larger lateral restraint to avoid detrimental bending along short wavelengths. So, from this point of view (although somewhat speculative), not only accounting for transverse load effect, a very small void fraction seems not beneficial for axial thermal contraction (WUCD) and 30 % is selected instead of the 27 – 28 % value from the TFPRO2-OST2.

The apparent axial flexibility for cables with longer twist pitches is not comprehensively understood or quantified up to now and only an intuitive concept of thinking can be offered at this stage of the design. For a better and more quantitative explanation detailed mechanical descriptions are required on at least the level as presented in [67][68][69]. For example in [68], all basic strain components for different kinds of wires were calculated to investigate the influence of the composite wire layout on the strain state of single wires. Eventually the efforts on strands needs to be implemented in a cable model as presented in [67], where a wide range of numerical simulations using several types of finite element models has been presented showing some analytical estimations for stretching and twisting of strand bundles. Also the work presented in [69] is already providing quantitative results that seem to add already to a better understanding of the average strain state and the strain distribution of the strands in a CICC.

However, the postulated hypothesis may not be easy to verify by quantitative modeling as even detailed cable models like presented in [69] may not be capable to fully catch the related mechanical phenomena in simulations, partly due to uncertainties in the input parameters but mainly due to size and complexity of the problem. The models are limited in the length and size and taking into account sufficient length, including the interaction between the conduit and all petals of the strand bundle would very likely be a requirement for appropriate modeling. However, in spite of lacking of a quantitative understanding of the mechanism behind axial flexibility, the experimental tests are the results needed for ITER. This cable axial flexibility may in particular be important for the CS conductor because the amount of steel in the cross section is larger than for the TF conductor (see Figure 10).

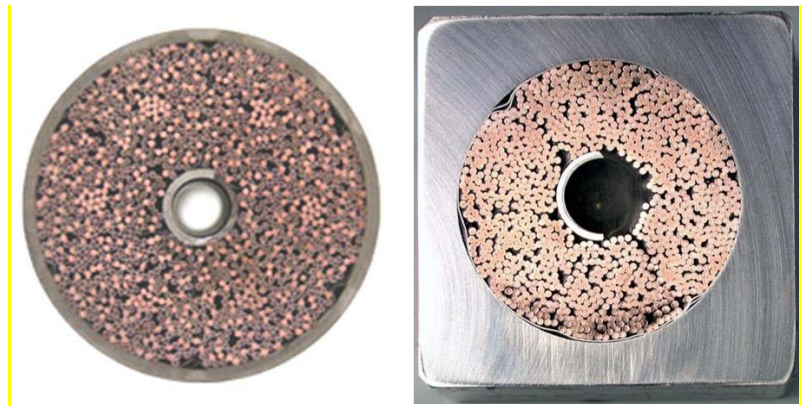

*Figure 10. The cross sections of the TF (left) and CS (right) conductors illustrating the large difference in the amount of steel in the cross section.*

In conclusion to the convincing experimental evidence of the TFPRO2-OST2 results, the analysis above supports the hypothesis that the performance of a CICC with long twist lengths is superior with respect to short and intermediate twist lengths, not only because of the high resistance against local strand deformations, but also because the Nb$_3$Sn layers are at less compressive effective axial strain already after cool down, leading to a higher current sharing temperature, and low compressive axial strain (see TFPRO2-OST2 results [25], [27], [34]). The increase of the shortest twist pitch from Option 1 (45 mm) to TFPRO2-OST2 (117 mm), lead to a gain in the effective axial strain of about 0.3 %, see Figure 8. This lead to a better performance for transverse load (no *IxB* sensitivity) and also reduced the WUCD effect, not only for short sample testing but also for the conductors in the ITER coils. As pointed out above, there is a strong *IxB* dependency of the performance observed on all ITER type TF and CS Sultan samples and the CS and TF Model and Insert Coils [7], [12], and WUCD effect as far as has been tested, except for the TFPRO2-OST2. This is well supported by the PITSAM results also showing better performance for electromagnetic load cycling and WUCD for longer twist pitches [27]. Next is then to find a cable pattern with not only generous strand support but also low interstrand coupling loss.

## 4 Coupling loss and cable patterns

### 4.1 JackPot-ACDC cable model and validation

Predicting the amount of coupling loss reduction that can be achieved by optimizing the cable twist pattern is practically hopeless without having a detailed electromagnetic cable model. Contrary to simple one stage configurations e.g. Rutherford type cables [70], the contact locations between strands in a CICC and the coupling with magnetic flux are much more complicated. For this, not only the trajectories of all strands in the cable are required with the strand to strand contact areas, but also their mutual inductive coupling. The only way out is the utilisation of a dedicated cable model incorporating all strand trajectories, their inductive coupling and associated current paths.

To achieve that, the numerical cable model JackPot-ACDC that contains a cable routine that calculates the trajectories of all strands in any CICC (>1000 strands), has been upgraded to also calculate interstrand coupling loss [32]. The model is capable of handling the strand scaling $I_c(B,T,\varepsilon)$, saturation, shielding, applied and self-magnetic field. In addition to the known parameters, such as strand- and cable diameter, void fraction and twist pitch sequence, it only requires one interstrand resistivity parameter. This parameter is obtained from direct measurements of the resistance between different pairs of strands in CICCs, obtained at 4.2 K and various loading conditions [20], [48], [49], [50]. The contact resistances and their associated distribution are calculated from the contact areas between the interstrand contacts (that depend straightforward on their trajectories) and the resistivity parameter.

Coupling loss measurements of different CICCs that are tested in the Twente Cable Press are used as benchmark for JackPot-ACDC's verification. One first example of an ITER PF NbTi conductor is given in [32], where the interstrand coupling loss was predicted by JackPot-ACDC based on the measured interstrand contact resistance distribution and the cable layout including dimensions and twist pitches of all stages. Another example is given in Figure 11 where the results of the measured interstrand contact resistance and the simulated results from an ITER sample (TFJA5-J) are displayed. In Figure 12 the measured total AC loss versus frequency is presented together with the total AC loss but with the hysteresis loss subtracted, compared to the predicted coupling loss by JackPot-ACDC. The prediction and measurement show good agreement although in general some deviations can be expected due to for example variations in twistpitch.

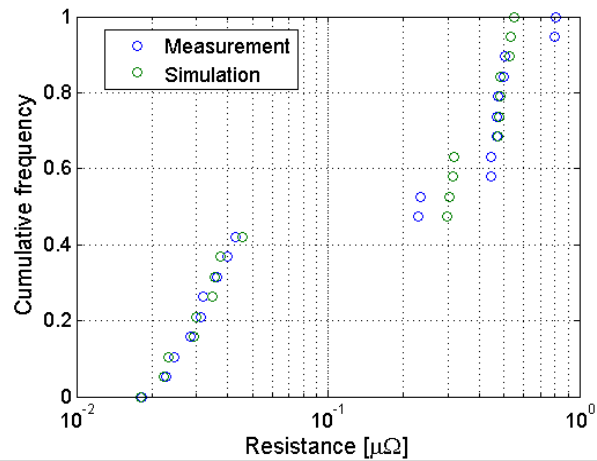

Figure 11. Comparison between the measured inter-strand and -bundle contact resistances on the TFJA5-J sample in the Twente Press and prediction of the JackPot-ACDC model based on a match of the contact resistvity parameter.

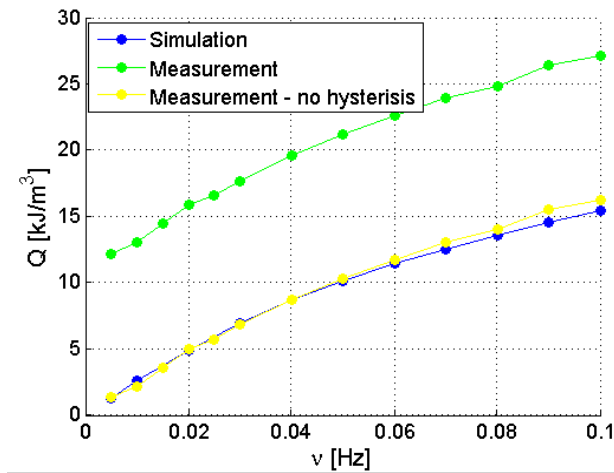

Figure 12. A validation of the JackPot-ACDC model based on Twente Press measurements of coupling loss and prediction based on the inter-strand and -bundle contact resistance measurements. After subtraction of the hysteresis loss, the prediction and experimental result are in good agreement.

Compression steps are performed in order to reach the correct trajectories of the strands after cabling. The final shape of the strand bundle depends on the choice of the compression algorithm. Another CICC model is used for comparison of the cable pattern after compaction, showing only minor influence on the coupling loss [71]. It appears up to now that for simulation of the evolution of the coupling loss with cycling in the press, a deviation can occur of about a factor of two at high bumber of cycles.

It is demonstrated that the amount of critical current degradation from transverse load [60], [61] is subject to interference due to different sub-cable twist pitches, we expect a similar behavior for coupling loss. This opens scope for optimisation of the twist pitch sequence for CICCs to minimise the transverse load degradation, the warm-up cool-down degradation and the coupling loss. At first sight, minimization of the coupling loss seems in conflict with reduction of the transverse load effect as longer twist pitches are needed for this. However, for reduction of the coupling loss the criterion should not just be the use of short pitches but to find the minimum area of the loops from linked strands and bundles which are coupled with the changing magnetic fields.

Several case studies on coupling loss are reported in this paper, starting with sinusoidal applied transverse fields with amplitudes that are commonly used in our AC loss test set ups. The hysteresis and intrastrand coupling loss, DC transport current and possible influence of joints are not included in the present analyses.

Most simulations are carried out using an interstrand resistivity (intra-petal) parameter $\rho_{ss}$ of $30 \cdot 10^{-6}$ $\mu\Omega m^2$. This value has been obtained as the average of the measured interstrand resistance distribution in three recent TF conductors tested in the Twente Press [48], [49]. The measured interstrand contact resistance between two strands after several hundreds of cycles amounts to 100 n$\Omega$m when in the petal, while it is 300 n$\Omega$m for strands from different petals. For the virgin condition these values are lower. The interstrand contact resistivity between strands from different petals ($\rho_{ip}$) is taken $1000 \cdot 10^{-6}$ $\mu\Omega m^2$. All simulations for an exponential dump presented further on were performed for an entire six petal cable and inter-petal resistance as mentioned above. The following simulations for the loss versus frequency were done for a single petal. The choice of simulating only one petal is dictated by the fact that the calculation of the strand mutual coupling is a heavy computational process, requiring long simulation time for parametric variations. For the initial parametric study, the assumption is justified that the main coupling loss is due to coupling currents between strands inside the same petal. In the single petal simulations, first the full cable with 5 cable stages according to the given cabling sequence and compression was build, then the AC field was applied to only one petal. It is well known that the classical single $n\tau$ value strongly depends on the selected frequency range. For this reason, the $n\tau$ evaluation is used for second order comparisons and the attention has been turned to the loss absolute value versus frequency.

### 4.2 Influence of cable twist pitch on coupling loss: sinusoidal applied fields

The simulations are performed for a sample length of 1 m exposed to a sinusoidal background field of ± 0.15 T and a frequency range from 0.005-2 Hz. The applied field values correspond to the test condition employed during the test of full-size ITER samples in the Twente Cable Press [20], [21], [42], [48], [49], [50], [51], [52]. The loss versus frequency for variation of the first stage twist pitch are shown in Figure 13 for a cable layout with 45-87-126-245-460 mm and in Figure 14 with 131-161-241-296-453 mm twist sequence representing the twist pitch lengths of the cable stages. The resistivity parameter for these cases is 5e-7 $\mu\Omega m^2$, taken from a fit of intra-petal resistance measurements in the virgin state of EUTF3 EAS sample [48]. The coloured bar at right in the Figures represents the percentage of twist pitch elongation. Both examples show that an increase of only the first stage twist pitch can already give a major decrease in overall coupling loss. For the same cable with $L_{p1}$=45 mm, it is shown in Figure 15 that a relatively small decrease of the $4^{th}$ stage twist pitch, leads to a significant decrease in the coupling loss. After an extensive parametric study, it appeared that in particular the ratio in the twist pitch sequence plays an important role and this can be understood as follows. From a JackPot-ACDC animation of the strand trajectories inside a CICC it is observed that with a traditional twist pitch sequence (as CS baseline) the radial distance between a pair of strands forming a coupling current loop varies largely along the cable, thus creating broad loops. Moreover, the radial position of a couple of strands changes as well and they move from the outer to the inner edges and vice versa as they rotate around the axis of the CICC along its length. When the twist pitch sequence is changed such as that the ratio between successive stages is kept closer to 1, the relative position of the strands is more constant and their relative distance varies less along their trajectories in the cable.

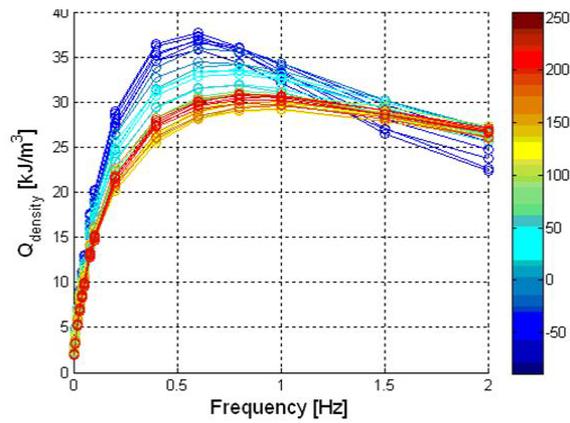

*Figure 13. The loss energy density vs frequency for variations of the first stage twist pitch of a cable with 45-87-126-245-460 sequence.*

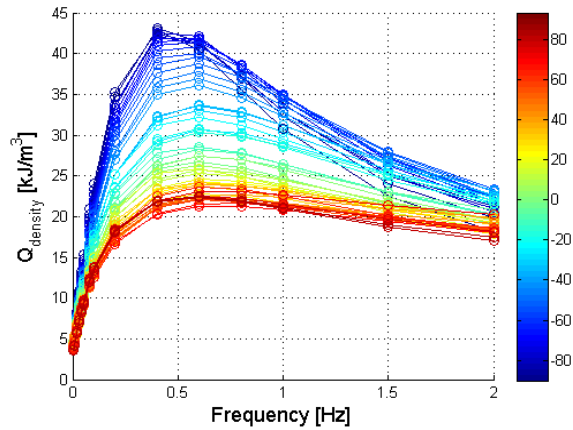

*Figure 14. The loss energy density vs frequency for variations of the first stage twist pitch of a cable with 131-161-241-296-453 sequence.*

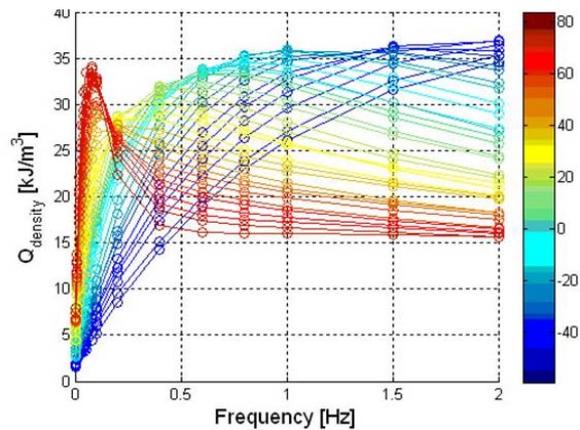

*Figure 15. The loss energy density vs frequency for variations of the fourth stage twist pitch of a cable with 45-87-126-245-460 twist sequence. A relatively small decrease of the $4^{th}$ stage twist pitch gives a significant decrease in coupling loss.*

The twist pitch of the last stage ($L_{p5}$, composed of the petals wrapped with steel tape) is typically fixed at 450 mm. The rationale behind a twist length of 450 mm seems more or less historically determined due to the limitation of the joint length which has to cover all petals although there is no explicit restriction to revise the last stage twist pitch. However, a larger last stage twist pitch length is not considered. When increasing the twist pitches of the initial stages to reach good mechanical stability, the decrease of the 4[th] and 5[th] (last stage) twist lengths becomes essential. This is well illustrated in Figure 15, leading to the key solution for low coupling loss in combination with larger twist pitches.

The ratio between the sequences in twist lengths of successive stages seems essential to achieve low coupling loss. So if starting from a short first triplet twist pitch ($L_{p1}$), a short last stage pitch seems required and vice versa. For the petal configuration the situation is slightly different but this will be discussed further on. More details and explanation can be found in [72].

Examples of the AC loss versus frequency for twist pitches having defined ratios between successive stages and with the first stage and the final stage twist pitch fixed at a chosen length is shown in Figure 16 and Figure 17. The first stage twist pitch $L_{p1}$ is selected as 80 mm, 100 mm, 120 mm, 150 mm, 200 mm, 250 mm and 300 mm, which means that for the larger twist pitches the higher stage sub-bundle pitches exceed the length of the petal twist pitch. The twist pitch ratio $\beta$ defines the cabling sequence from one cable stage to the next as: $L_{p2}= \beta*L_{p1}$, $L_{p3}= \beta*L_{p2}$ etc., $\beta$ is varied from 1.05 to 1.50.

From the simulation results we find an increase of the loss with larger first stage twist length and with larger stage ratio. However, combinations can be found that produce significantly lower AC loss than the present ITER CS baseline design but with much longer twist pitch length of the initial stages.

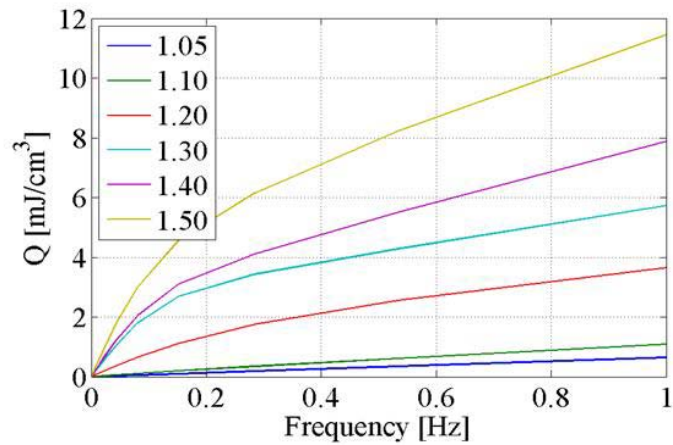

*Figure 16. The AC loss versus frequency for twist pitches having defined ratios between successive stages (the first stage twist pitch is fixed at 100 mm and twist pitch 5 (petals) is fixed at 450 mm).*

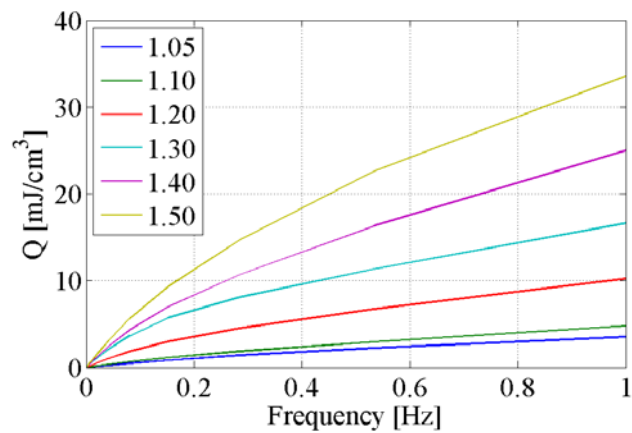

*Figure 17. The AC loss versus frequency for twist pitches having defined ratios between successive stages (the first stage twist pitch is fixed at 200 mm and twist pitch 5 (petals) is fixed at 300 mm).*

A summary of the coupling loss time constant $n\tau$ values representing the initial slope of the AC loss versus frequency curves is given in Figure 18 for all combinations. It appears that the intra-petal $n\tau$ is not systematically largely affected by the last stage petal pitch. The main message from this plot is that the ratio $\beta$ must be kept close to 1 for low coupling loss.

Based on the coupling loss results from above, but also on the study on strand support from below, we selected three alternative CS cable layouts for some further comparisons. One layout is chosen for low AC loss and large lateral strand support CS-Twente, the second layout CS-S1 is chosen for its high stiffness and relatively low loss while the third is chosen because it has similar AC loss as the CS-Baseline for the higher frequency range and very favorable lateral strand support. The layouts are listed in Table 3.

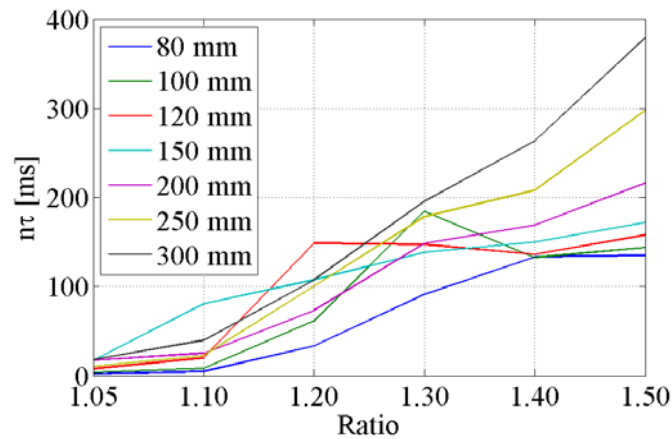

*Figure 18. The $n\tau$ versus twist pitch ratio between successive stages for different first stage twist pitch length. The last stage twist pitch (petals) is fixed at 300 mm.*

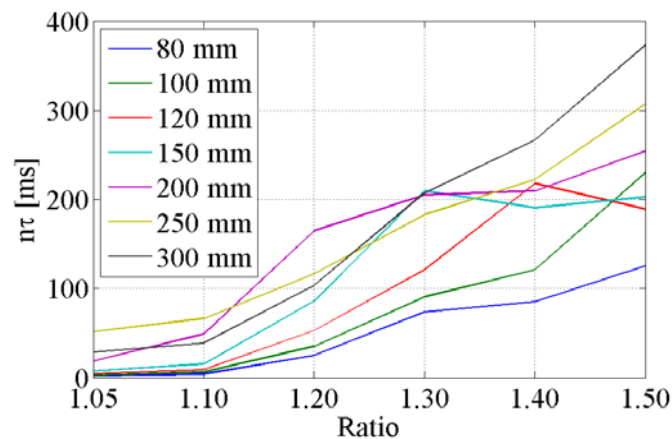

*Figure 19. The $n\tau$ versus twist pitch ratio between successive stages for different first stage twist pitch length. The last stage twist pitch (petals) is fixed at 450 mm.*

Table 3. Alternative CS cable layout specifications based on minimization of coupling loss and high lateral strand support. The conductors are cabled according to the (2sc + 1Cu) x 3 x 4 x 4 x 6 configuration. The CS-A layout is optimized for very low AC loss, while the CS-S1 is optimized for large lateral strand support and accordingly low strain degradation. The CS-S2 layout is for comparison.

| Twist pitches [mm] | CS-Twente very low AC loss good mechanical stiffness | CS-S1 high mechanical stiffness low AC loss | CS-S2 high mechanical stiffness high AC loss |
|---|---|---|---|

| Lp-ratio | 1.1 | 1.1 | 1.2 |
|---|---|---|---|
| Lp1 | 110 | 200 | 200 |
| Lp2 | 118 | 220 | 240 |
| Lp3 | 126 | 242 | 280 |
| Lp4 | 140 | 266 | 346 |
| Lp5 (petal) | 350 | 300 | 450 |
| Optional Petal coverage | 70 % | 70 % | 70 % |
| Void fraction | 30 % | 30 % | 30 % |

The $n\tau$ values calculated from the initial slope amounts to 58 ms for the CS-Baseline, 18 ms for the CS-Twente, 25 ms for the CS-S1 and 156 ms for the CS-S2. It should be noted that $n\tau$ from the initial slope is a conservative approach. A lower void fraction as compared to the CS-baseline, also enhancing the radial constraint, is allowed because of the lower coupling loss generation.

In Figure 20 a comparison is made for coupling loss versus frequency for different cabling twist pitch lengths from short to very long twist pitch. The cable patterns are compared for the CS-baseline design, the IO short pitches proposal, (see Table 1 for details, the TFPRO2-OST2 long pitches pattern and the present TF-Option2 pattern and the three Twente CS patterns from Table 3. In addition a cable simulation example with 305-335-370-410-450 mm (stage ratio=1.1) is included to demonstrate that the combination of very long pitches and low loss (lower than CS baseline) is possible with extremely large pitches.

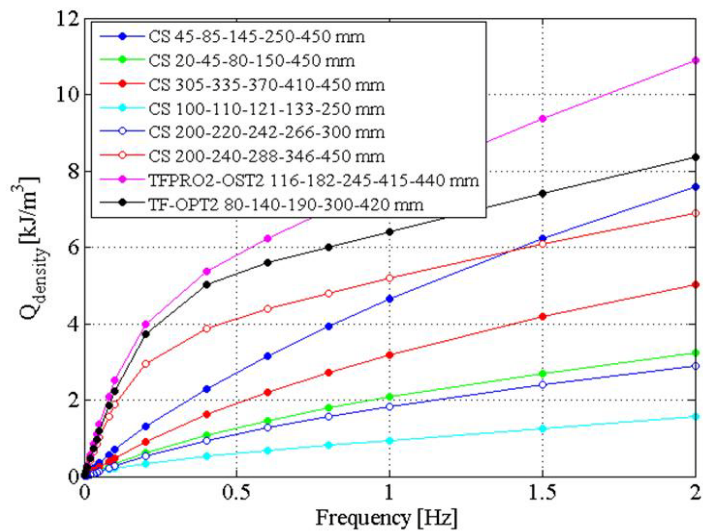

*Figure 20. AC loss versus frequency for varied twist pitch sequences, illustrating the relevance of short and long pitches for frequencies up to 2 Hz.*

The TF Option 2 cabling leads to comparable amount of AC loss generation as the CS baseline layout for higher frequencies but for low frequencies, most relevant for ITER, the AC loss differs substantially. The TFPRO2-OST2 sample shows highest AC loss, which is in agreement with the SULTAN test result [24] but below 0.2 Hz, the AC loss is practically similar to TF-Option2. The CS-alternative with very short pitches, starting at 20 mm for the first triplet has low AC loss. However, important to notice is that the layout with very large twist pitches starting at 305 mm for the first triplet generates loss that remains restricted to a level below that of the CS baseline, comparable to that of the IO CS alternative with very short pitches. For industrial scale cable manufacture, the long twist scheme sequence is not considered as largely problematic and it may even be possible that this cabling process can be well controlled [73]. However, we should be cautious here as we have not evaluated so far how close the cable compaction in the JackPot-ACDC model follows the probability of strands getting disengaged from their triplets and forming larger area loops then what is identified in the model. This may eventually be a limitation for application of very long pitches like the $L_{p1}$=200 mm example.

The cable having the same twist pitch sequence but without copper strand in the triplet (three superconducting strands in the first triplet with Cu:nonCu=1.5 instead of 1.0) has AC loss only slightly higher than that of the CS baseline layout. This implies that the $n\tau$ value calculated from the initial slope is still close to the 75 ms criterion for the CS conductor [40]. The requirement which is set on the cabling pattern must be considered for the relevant frequency (dB/dt) range.

### 4.3 Coupling mechanism and local power dissipation

Besides the loss density, also the local distribution of the coupling loss power inside a cable for differences in the cable pattern can be relevant. Although more detailed studies are ongoing, some preliminary results are discussed here. The strand current cumulative distribution for a sinusoidal applied firld at 100 mHz for full sample lengths of 1 m and 10 m is depicted in Figure 21, confirming that the cable length in the simulation has hardly any influence on the results. The results are shown for the CS baseline, the short pitch and the CS-Twente cable layout.

When comparing the maximum strand currents which are induced in a CS baseline design and the extreme case of for example a long 305-335-370-410-450 mm twist sequence, there are a few strands that carry higher coupling current in the cable with larger pitches. The majority of the strands in the cables with longer pitches follow practically the same current distribution as in the CS baseline cable and only a very small fraction of the strands reaches currents of maximal a factor of two higher. Only the CS-S2 option has a significantly larger spread and is also reaching much higher currents.

The AC strand currents in the 305-335-370-410-450 mm cable appear to vary less along the direction of the cable axis (some of them stay constant along significant portions of the cable length. This is likely connected to the lower number of contacts between the strands per unit length.

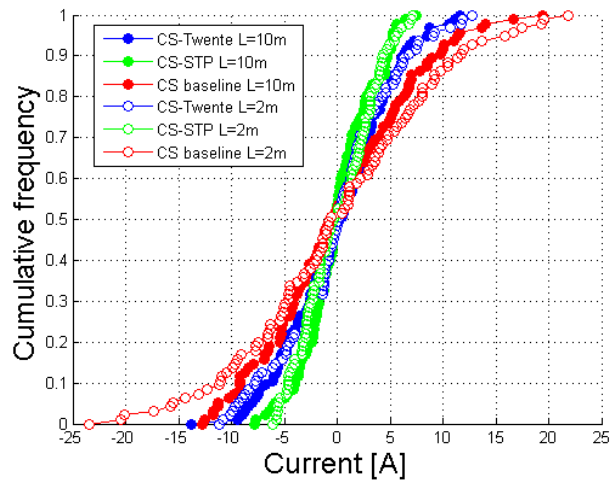

*Figure 21 Strand current cumulative distribution (at applied field Ba=400 mT and f=100 mHz) for full sample lengths of 1 m and 10 m in order to find possible influence of the cable length in the simulation.*

Knowing that a fraction of the strands that carry a higher coupling current, it is interesting to compare the AC power loss density for different cable length as presented in Figure 22. Also here it is confirmed that the cable length in the simulation has hardly any influence on the results.

Other results showed that the TF-Option2 and TFPRO2-OST2 layout produce appreciably higher power with substantial increase towards the ends of the conductor. This end-effect is also noticed for the CS-S2 layout. The twist ratio of the TFPRO2-OST2 and the TF-Option2, although not constant, is about 1.5 and together with the longer pitches this seems to explain the high coupling loss. Also for the CS baseline, although having shorter pitches, the twist ratio leads to relatively high loss.

During cable manufacture some variations in the twist pitch are unavoidable and we looked at the effect on the coupling loss versus frequency by varying only the 4[th] stage twist pitch, the result is depicted in Figure 23. For the first three cable stages the effect is much less, in particular at the lower frequencies.

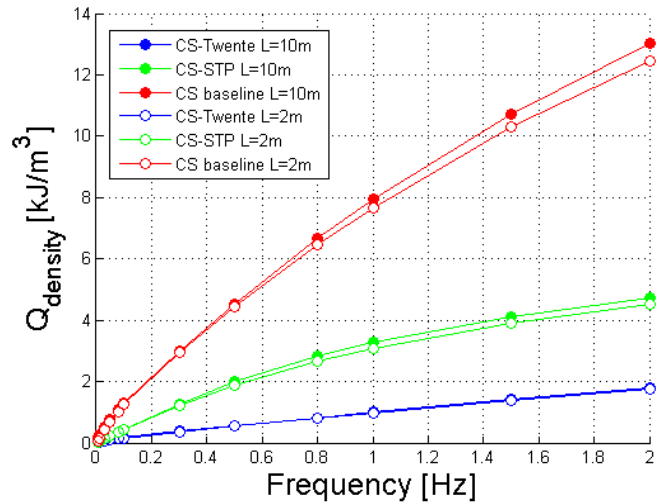

*Figure 22 AC power loss density versus frequency for full sample lengths of 1 m and 10 m in order to find possible influence of the cable length in the simulation (Ba=400 mT).*

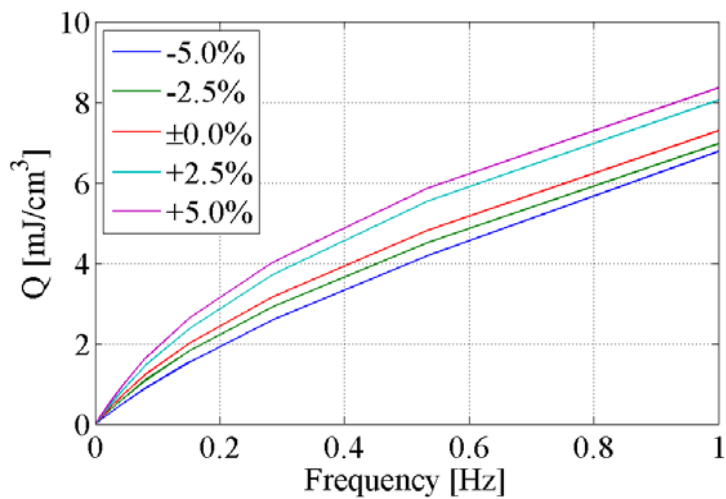

*Figure 23 AC power loss density versus frequency for full small varion in the $4^{th}$ stage twist pitch length of the CS-Twente cable layout.*

**5 Strand to strand contact length**

The combination of cabling sequence ratio and first stage twist pitch that leads to mechanically good performance for $Nb_3Sn$ strands in a CICC, is evaluated by analysing the average strand to strand contact length. This parameter can be assessed with JackPot-ACDC, following all the strand trajectories and interstrand contacts, including the copper strands when present in the triplet. The basic hypothesis is that for longer contact lengths between strands, the distribution of loads is more homogeneous and as such providing good lateral strand support, less strain alteration and thus improved mechanical stability. It is obvious that for NbTi cables, not suffering from strain sensitivity, twisting scheme's can be applied with the focus only on low AC loss.

The average contact length has been obtained summing the length of each contact of a strand with all the others and dividing it by the number of contacts. This is done for all strands in the cable. Besides the average contact length it is also interesting to know the distribution of the contact lengths in the cable and this is shown for the CS baseline conductor layout, four alternative CS layouts and the TFPRO2-OST2 in Figure 24.

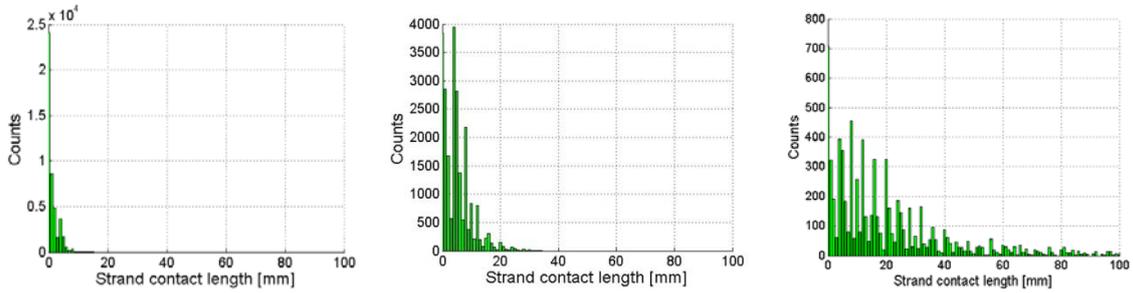

*Figure 24. The distribution of the number of strand to strand contacts versus the contact length for the CS alternative with $L_{p1}$= 20 mm in the first triple (left), CS baseline conductor layout (middle) and CS-Twente with $L_{p1}$=110 mm and ratio $\beta$=1.1 (right),.*

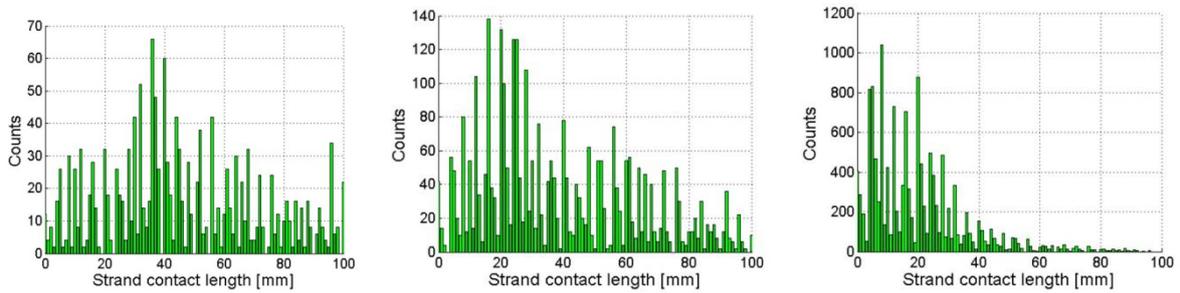

*Figure 25. The distribution of the number of strand to strand contacts versus the contact length for the CS-S1 with $L_{p1}$=200 mm and ratio 1.1 (left), CS-S2 with $L_{p1}$=200 mm and ratio $\beta$=1.2 (middle) and TFPRO2-OST2 (right).*

The contact length for the CS baseline has pronounced peaks in the number of counts around 2 mm and at 5 mm. The CS-Twente already has a more homogeneous distribution with high peaks up to 20 mm The alternative design for very high mechanical support and modest coupling loss CS-S1, indeed shows a much more homogeneous distribution with its peak around 40 mm. For the CS-S2 design demonstrating the effect of the increased ratio β=1.2, the peak is at 20 mm, apparently due to the higher $\beta$ and the TFPRO2-OST2 distribution is more or less comparable to that of the CS-Twente.

The average contact length for the various cable layouts is summarised in Figure 26. We find that, when adding twice the strand diameter to the average contact distance computed by JackPot-ACDC, the characteristic bending wavelength as defined for the TEMLOP predictions is in nice agreement, but this time with improved fundamental detail and accuracy. For the ITER Option1 cabling pattern we anticipated $L_w$ to be 7 mm, for the TFPRO2 long pitches we used 26 mm and for the Option2 layout an $L_w$=16 mm was taken [23], [25]. The average contact length for the alternative design with short pitches ($L_{p1}$=20 mm) is practically similar to that from the CS baseline and TF-Option1; 5 to 6 mm. The average contact length for the TF Option 2 layout is 13 mm, while for the TFPRO2-OST2 it amounts to 19 mm and for the CS-Twente design it is 26 mm. Notice that the contact length of the CS-S1 and CS-S2 layouts are 86 mm and 51 mm respectively, indicating that a higher $\beta$ may cause a decrease of the average contact length.

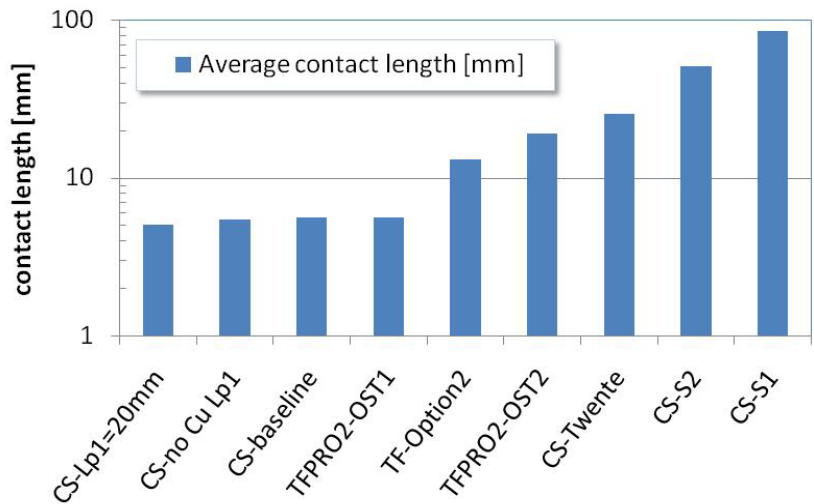

*Figure 26. The average strand to strand contact length for various cable layouts (logarithmic scale).*

The average contact length for cables with different first stage twist pitch and varying ratio $\beta$ in the twist sequence from 1.05 to 1.50 and a last stage twist pich of 450 mm is presented in Figure 27. The average contact length does not vary a lot with the ratio for the shorter first stage twist pitches of 80 mm to 150 mm, although we observe the tendency to a maximum at a ratio of 1.1. However, already at $L_{p1}$=80 mm the average contact length has reached the level of the TFPRO2-OST2 long pitch conductor of almost 20 mm. The cable with $L_{w1}$=120 mm and ratio 1.5 may have more or less similar properties as the TFPRO2-OST2. A change of the last stage (petal) twist pitch is investigated as we belive this is attractive for increase of cable axial flexibility in view of the WUCD effect. The results for the last stage pitch fixed at 300 mm are depicted in Figure 28, illustrating that a decrease of the last stage twist pitch enhances the average strand to strand contact length, in particular for larger $L_{w1}$ and smaller $\beta$. Computations of the interstrand coupling loss however revieled that going to shorter last stage pitch increases the loss again.

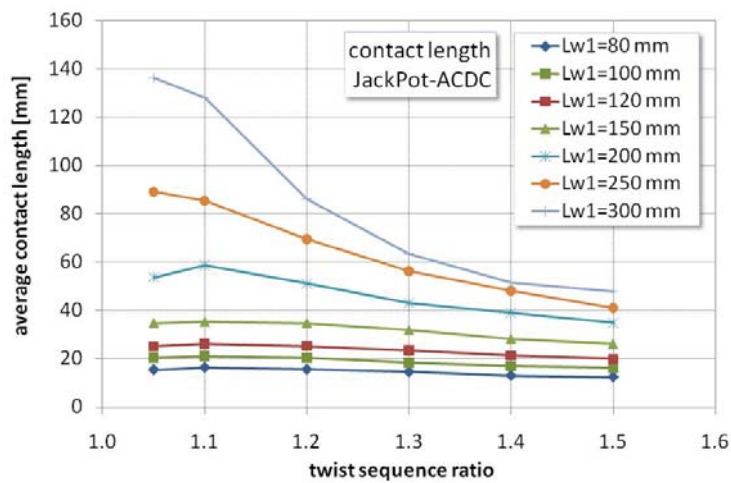

*Figure 27. The average strand to strand contact length versus the twist pitch sequence ratio for different first stage triplet twist pitch, including copper strands and with the last stage pitch is fixed at 450 mm.*

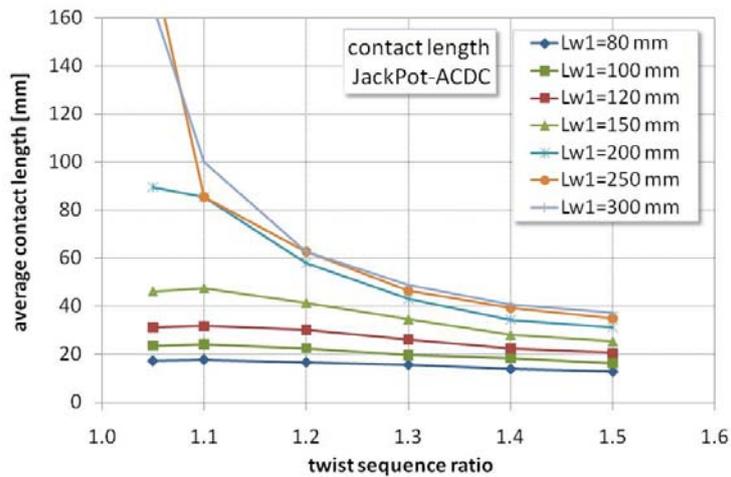

*Figure 28. The average strand to strand contact length versus the twist pitch sequence ratio for different first stage triplet twist pitch, including copper strands and with the last stage pitch is fixed at 300 mm.*

For a reduction of the AC loss, we know now that the pitch ratio must be close to 1. Although perhaps not practical from cable manufacturing point of view, we also investigated the effect of ratio's smaller than 1, a ratio varying with cable stage and the mixing of right- and left hand twisting but we did not find any favourable combinations..

In conclusion, we discovered that reducing the ratio $\beta$ is not only beneficial for the AC loss but also for the increase of the average contact length. An example of the structural difference in cabling pattern with CS baseline and alternative option CS-Twente and CS-S1 is given in Figure 29. It shows the full CICC including all five stages and it is clearly visible that for the CS-Twente and to even larger extent for the CS-S1, the strands are much better supported by surrounding strands providing lateral constraint against local deformations. It can be easily understood that for the CS baseline layout or shorter pitches the opportunity for buckling in case of axial cable contraction is larger.

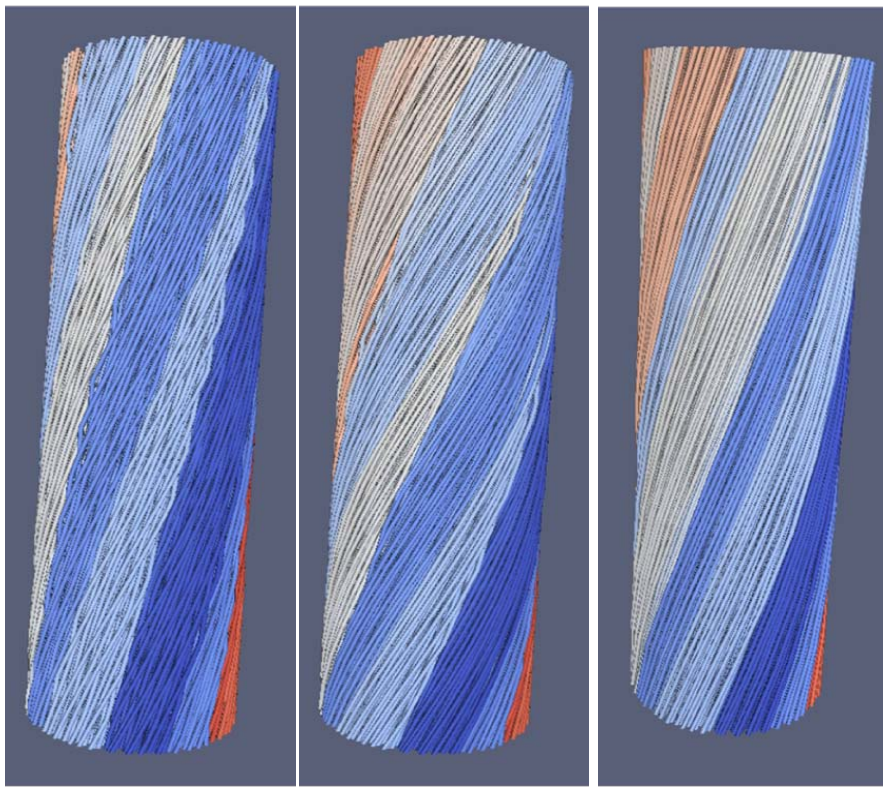

*Figure 29. The cabling pattern of the CS baseline (left), CS-Twente (middle) and CS-S1 (right).*

## 6. Bending strain calculated with CORD

We assume that, although the mechanism of axial thermal contraction is different, the simulation of an axial tensile strain test to some extent reflects the occurance of bending strains in a cables conductor. The average bending strain for an applied axial tensile strain of 0.6 % on a cable, calculated with the mechanical cable model CORD (only first three cabling stages) for several twist pitch combinations, is represented in Figure 30 [60], [61]. The bending strain of the CS-Twente layout with 110 mm first stage pitch is small compared to that of the CS baseline but also much lower than what is found for the alternative design with very short pitches. Obviously the combinations with larger $L_{p1}$ and $\beta$ like the CS-S2 layout with 200 mm first stage pitch tend to have very low bending strain.

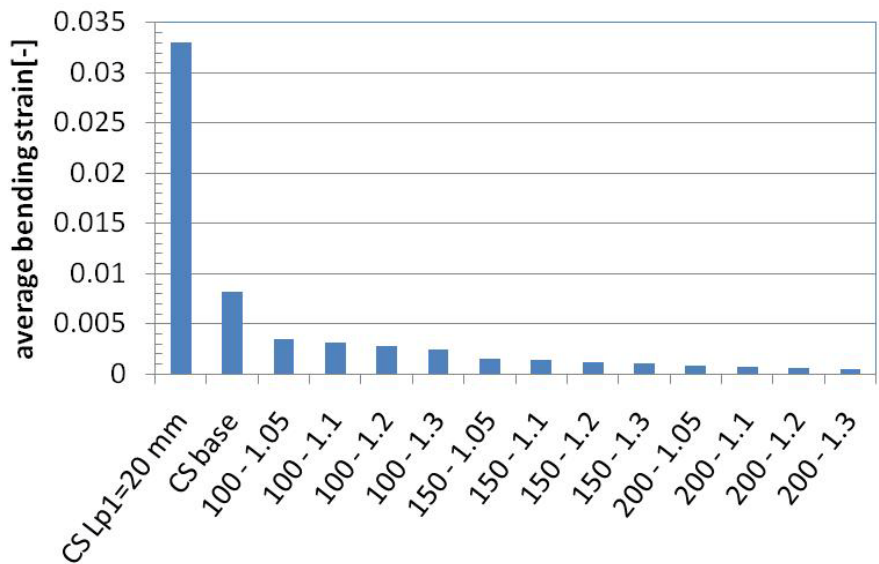

*Figure 30. The average bending strain for an axial tensile strain of 0.6 % on a cable, calculated with CORD (only first three cabling stages) for several first stage twist pitch and ratio combinations.*

## 7 Coupling loss and field exponential dump

In the parametric coupling loss study from above, we have used sinusoidal swept fields. Here the case is considered of exponential decrease of the ITER CS conductor current with a time constant of 7.5 s (design approach) approximating the scenario that the CS conductor current decrease is driven by the simultaneous dump of the coupled CS and PF system from (End Of Burn) EOB conditions as defined in the 15 MA scenario V1.10 [28], [40]. A safety factor of 2 on the calculated electromagnetic energy loss from a disruption should be allowed, to cover both uncertainties in the calculation and the impact of transport current on the AC losses. According to the ITER MDDD [40], there is a very large margin for all the coils for all conditions under analysis, with the exception of the CS. Using the disruption data in section 5.2 of MDDD11-1, the heat input to the CS conductor is (from 0.25 s to 2 s) 308 mJ/cm$^3$. The effective 'single value' time constant for the AC loss assessment for the CS cable: $n\tau dB/dt$=0, $N$=100, "full load") =75 ms ($N$ is the number of load cycles) as the $n\tau$ amounts to 200 ms for the TF conductor.

For the computation of the exponential dump, three different twist pitch sequences were taken apart from the CS baseline from Table 1, From the same Table, the alternative CS design with three superconducting strands in the triplet, the layout with short twist pitches and the optimized pattern proposed as CS-Twente and listed in Table 3. The decay of the background field and strand current is depicted in Figure 31 for the CS baseline conductor. Also shown is the critical current of an internal tin strand [25] at this field, and at 4.5 K and -0.6 % axial strain as for this simulation no critical current limitation was used. The -0.6 % axial strain value is chosen rather arbitrarily in this example, as the analysis for TF conductors seems to point towards an average axial strain value of about -0.4 %, while for CS conductors with larger steel content in the cross section, a somewhat higher compaction is assumed. Ignoring the critical current saturation for this case was done to illustrate that for a fraction of the time, the currents exceed the critical current and the actual power dissipation during the intial seconds would have been less in reality. On one hand, this can be considered as conservative but on the other hand some extra margin is required because of the interaction between coupling current and transport current. The corresponding power dissipation versus time is presented in Figure 32 for all four conductor layouts. The power dissipation of the CS-Twente conductor is far below that from the CS baseline, while for the CS-short pitch conductor the power is in between that of the CS-Twente and the CS baseline. The difference between having a copper strand in the triplet or three superconducting strands appears to be marginal, although there is higher loss for the case with three superconducting strands.

In Figure 33 the AC loss is plot as a function of the frequency of an applied field with amplitude 0.15 T for all conductor layouts. The striped line represents a coupling loss time constant of 75 ms. For low frequencies the AC loss for the CS-baseline conductor with two superconducting strands but also the layout with three superconducting strands in the triplet, both stay at the criterion of $n\tau$=75 ms..

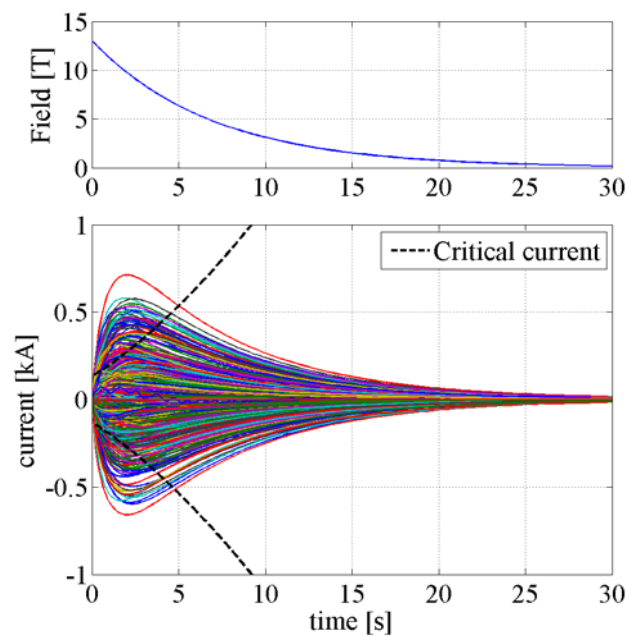

*Figure 31. Background field decay (top) and strand current (bottom) for the CS baseline conductor. Also shown is the critical current of an inter tin strand (as used for the TFPRO2-OST1) [25] at this field, and at 4.5 K and -0.6 % axial strain.*

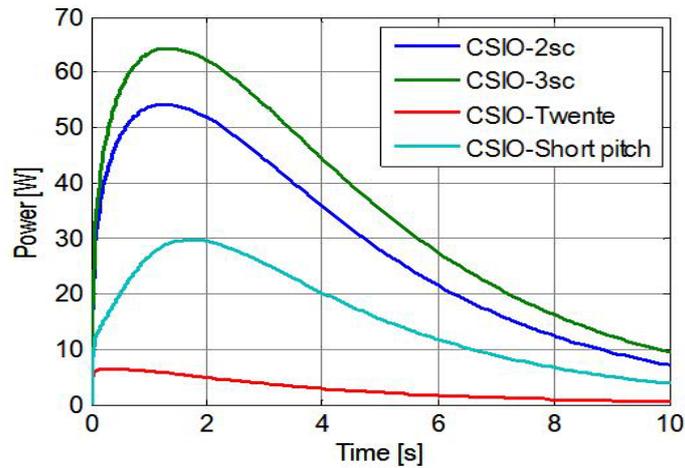

*Figure 32. Accumulated energy dissipation compared for four conductor layouts, showing that the peak power of the CS-Twente conductor layout is an order of magnitude reduced compared to that of the CS baseline.*

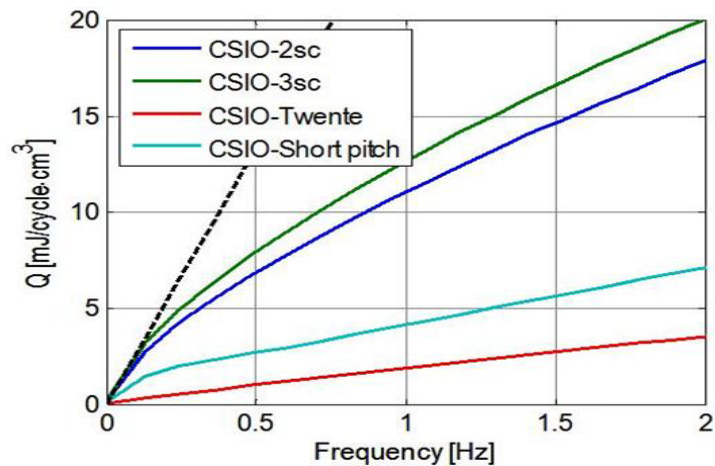

*Figure 33. The coupling loss as a function of the frequency of a sinusoidal applied AC field. The amplitude of the applied field is 0.15 T. The striped line represents an $n\tau$ of 75 ms.*

**8 Recommendations**

With the use of the developed models, an optimum solution is predicted satisfying the requirements for both AC coupling loss and mechanical stability for the ITER CS conductor. At first, the prediction should be verified experimentally by testing a short sample (in Sultan). The twist pitch off all stages, including the last stage and the petal wrap coverage is optimised for the CS-Twente as listed in Table 3. The cable is specially designed for minimum AC loss and moderate strand lateral restraint. The alternative layout with $L_{p1}$=200 mm is presented to demonstrate the scope for further optimised strand support (CS-S1) and still small AC loss, lower than the present ITER CS baseline layout.

We propose a void fraction of 30 % since the AC loss will be low and less helium is required to absorb the heat during fast field transients.

Additional benefit may be expected from 70 % wrap coverage instead of 50 % in preventing disengagement of the strand bundles and maintaining good strand lateral restraint against local deflections. Larger petal wrap coverage could help in preventing the initiation of kinks in case of relatively large magnet field gradients along the conductor length, in particular at the gap between cable and jacket. The required ability for current redistribution, needed to adjust for the non-uniformity introduced by joints, is not pre-dominantly between petals but mainly within the petals.

In the meantime four conductor lengths have been manufactured according to the cable layouts as presented in Table 4. Photographs of the petals of the three cabling layouts, without and with steel wraps are shown in Figure 34. The Sultan samples are being manufactured according to the Sultan standard procedure as developed for the TF short sample testing, including crimping rings, several strain gauges along the sample length in high and low field region, voltage tap and temperature sensor arrays and fully soldered terminations for the joints.

Not only AC loss measurements are requested from the Sultan sample but in addition short sample AC loss tests, preferably with load cycling, should be performed in a different facility, being able to cover the relevant low frequency range. The reason for this is that in particular the low frequency range is important for the exponential dump associated with the simultaneous dump of the coupled CS and PF system.

After confirmation that the prediction leads to sufficient minimisation of the $T_{cs}$ cyclic degradation and AC loss compared to the present CS baseline conductor layout, the CS-ITER conductor design may be adjusted to the CS-Twente layout or a layout with further fine-tuning (if needed) in the desired direction of benefit, strain degradation or AC loss.

Table 4. Cable layout of the four CSIO alternatives to be tested in 2012.

| Twist pitches [mm] | CSIO-2sc baseline | CSIO-3sc | CS-Twente | CSIO-Short Twist Pitch |
|---|---|---|---|---|
| Lp-ratio | 1.8 | 1.8 | **1.1** | 2.0 |
| Lp1 | 45 | 45 | **110** | 20 |
| Lp2 | 83 | 83 | **118** | 44 |
| Lp3 | 141 | 141 | **126** | 78 |
| Lp4 | 252 | 252 | **140** | 156 |
| Lp5 (petal) | 423 | 423 | **352** | 423 |
| Petal coverage [%] | 70 | 70 | **70** | 70 |
| Void fraction [%] | 30 | 30 | **30** | 30 |

CS-Short Twist Pitches          CS baseline pattern          CS-Twente design

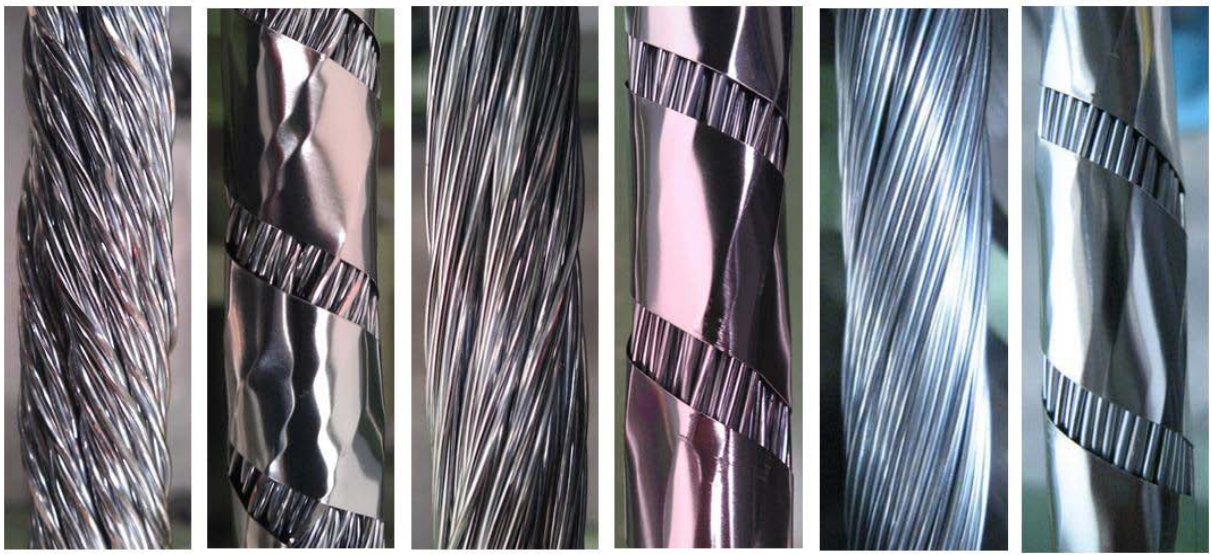

Figure 34. The petals after cabling for three different cable patterns; the CS short twistpitch, the CS baseline and the CS-Twente layout. The petals are presented in pairs, before and after wrapping the petal with steel tape.

**9 Conclusions**

We succeeded in developing detailed models and guidelines that can be used for the optimisation of $Nb_3Sn$ CICCs superconductors with particular reference to the ITER CS conductor. The optimization is focused on minimum coupling loss in combination with least possible sensitivity for degradation due to electromagnetic and thermal loads. The models provide new and valuable insights on the relation between interstrand coupling loss and the design of the cable pattern with complicated structures as used in CICCs.

In a multi-stage cabled CICC, long twist pitches do not necessary lead to copious increase of the coupling loss but can be well controlled. The most important conclusion of this paper is that the problem of the severe degradation of large CICCs, as designed for ITER, can be solved by increasing the lower stage cabling pitches and selecting a pitch sequence that reduces the interstrand coupling loss. This is reached when the first stage twist length is taken long (100 - 200 mm) and the ratio in the twist pitch sequence of subsequent stages is chosen just above 1.

Long twist pitches as a solution against transverse load degradation is based on a more homogeneous distribution of the stresses and strains in the cable, providing better strand to strand support and significantly moderate the local peak stresses associated with short twist pitches, most of all lateral restraint is required. This was already previously predicted by the TEMLOP model but with the JackPot-ACDC model, the average strand-to-strand contact length is accurately determined and applied as a valuable criterion for strand lateral support.

Also the mechanical response of the cable during axial contraction during cool down is improved. For short pitches periodic bending in different directions with relatively short wavelength is imposed due to lack of sufficient lateral restraint. This can lead to high bending strand and eventually buckling. Whereas for cables with long twist pitches the strands are only able to respond as bending bundles with sliding strands, being tightly supported by the surrounding strands, providing sufficient lateral restraint of radial pressure in combination with enough slippage to avoid single strand bending along detrimental short wavelengths. Together with a void fraction of 30 %, this results into axial flexibility with limited bending strain and torsion with minimum degradation. Important experimental evidence is already provided by the test on the TFPRO2-OST2 sample, which is still until today the best performing TF sample ever without any electromagnetic or warm up – cool down cyclic load degradation. The analysis shows that increasing the twist pitch can lead to a relaxation of the effective axial strain of more than 0.3 %.

A further improvement of the strand lateral support is larger petal wrap coverage of at least 70 %. This would help in preventing the initiation of buckling in the cable-jacket gap region initiated by the combination of transverse and axial load in case of relatively large magnet field gradients along the conductor length.

Following this approach, the design can be significantly improved with utmost minimum changes in the conductor design. We consider a design change as mandatory since the average degradation rate for the recently tested CS samples is too high and for the CSMC-1A it even reaches to 2.5 mK/cycle, revealing that any cable layout close to that of the CS baseline design will degrade.

We propose to first validate the prediction experimentally by a short sample test in Sultan on a sample following the CS-Twente cable layout. The second test is an AC loss test on a short section in a dedicated facility for the relevant low frequency range.

**UNIVERSITY OF TWENTE.**

**Acknowledgements**

This work is supported by the ITER International Organization, Cadarache, France; F4E, Barcelona, Spain; and the University of Twente, Enschede, Netherlands.